\documentclass
[twocolumn,showpacs,preprintnumbers,amsmath,amssymb]{revtex4}

\usepackage{graphicx}
\usepackage{dcolumn}
\usepackage{bm}
\usepackage{rotating}
\usepackage{color}
\usepackage{soul}
\usepackage{bigints}

\begin{document}
\title{Snell's law for the Schwarzschild black hole}
\date{\today}
\author{X. H. Zheng$^1$}
\email{xhz@qub.ac.uk} 
\author{J. X. Zheng$^2$} 
\affiliation{$^1$Department of Physics, Queen's University of Belfast, BT7 1NN, N.  Ireland}
\affiliation{$^2$Department of Electrical and Electronic Engineering, Imperial College London, SW7 2AZ, England}

\begin{abstract}
The Wheeler equation, for electromagnetic disturbances in a gravitational field, was found by Fiziev to have exact solutions both above and below the event horizon, in the form of waves propagating both inwardly and outwardly.  This observation can be interpreted and verified from the optical point of view, entirely on the basis of the Schwarzschild metric for length contraction and time dilation, in order to derive a differential version of Snell's law for the Schwarzschild black hole.  It reveals interesting physics, including the correct amount of light deflection by the Sun, internal and external Oppenheimer-Snyder cones of the black hole, properties of its phonon sphere and the conclusion that light-rays are kept below the horizon by length contraction and time dilation rather than deflection.
\end{abstract}
\maketitle

\section{Introduction}\label{sec:1}
Electromagnetic fields in the Schwarzschild metric have been investigated for a long time \cite{Bonnor, Lovelock, Papapetrou, Linet, Bachelot, Bachelot2, Gutsunaev, Choquet, Kroon, Blue, Ovsiyuk, Fernandes} and Fiziev secured an unusual advance in 2006 \cite{Fiziev}.  With an explicit substitution he transformed the Wheeler equation \cite{Wheeler} into Heun's equation \cite{Ronveaux} which is known to have exact solutions, both above and below the event horizon of a black hole, in the form of waves propagating both inwardly and outwardly.  He presented a number of such solutions and commented on their eventual physical significance and some open problems \cite{Fiziev}.

The Wheeler equation was derived in 1955, from a system of fundamental equations for coupled electromagnetic and gravitational fields, to study strong electromagnetic disturbances capable of distorting spacetime to trap themselves \cite{Wheeler}.  Sometimes the Wheeler equation is referred to as the Regge-Wheeler equation probably because, in 1957, the two authors jointly published a paper on a related topic \cite{Regge}.  In the case of a black hole and weak electromagnetic disturbances the Wheeler equation assumes a specific form in the so-called tortoise coordinate in association with the Schwarzschild metric.  With this specific Wheeler equation Fiziev applied his substitution and linked it to Heun's equation.

Heun's equation was studied by Karl Heun in the late 19th century and still is an active topic of contemporary investigation.  It is a second order ordinary differential equation with 4 singularities in its general form.  By the process of confluence the number of singularities can be reduced (to 3 in the case of the Wheeler equation) for a rich variety of useful and important applications.  Solutions to Heun's equation include local solutions (power series), Heun functions, Heun polynomials and path-multiplicative solutions \cite{Ronveaux}.  The solutions Fiziev found are mostly local solutions, where a countable set exists below the event horizon, simultaneously finite at the horizon and centre, and obviously square integrable.  In his opinion, if these are proven to form a complete set in the corresponding functional space, they will present a natural basis of normal modes for a well-defined expansion of any small perturbation of the metric in the interior of the black hole \cite{Fiziev}.

The perspective of Fiziev amounts to a sophisticated scheme to analyse waves, essentially via Fourier expansion in terms of the so-called normal modes, which are solutions to the Wheeler equation, with a boundary value (zero for example) imposed upon the event horizon by some physics which, to the best of our knowledge, has not yet been identified.

Here we tackle the problem from the optical point of view so that, as long as light-rays exist in the void below the event horizon, in addition to their presence above, it is sufficient for us to know length contraction and time dilation from the Schwarzschild metric to determine the speed of the waves in order to trace the rays, with no need to identify additional physics.  The result is a differential version of Snell's law which, like its counterpart in classical optics, reveals interesting physics.

For example, above the event horizon, our version of Snell's law enables us to find the correct amount of light deflection by the Sun via straightforward evaluation.  It enables us to define quantitatively the external Oppenheimer-Snyder cone, together with properties of the so-called phonon sphere, to determine whether or not a ray launched above the event horizon can escape from the gravitational field.  It also tells us clearly that, if incapable of escaping, the ray must encounter the event horizon at a right angle.

Below the event horizon, we find from our version of Snell's law the internal Oppenheimer-Snyder cone to determine the allowed angles for light-rays to propagate in both the outward and inward directions.  Historically Einstein started with optics to derive the general theory of relativity \cite{Einstein}, adding length contraction and time dilation to the arsenal of optics as devices to manipulate light-rays.  We conclude that  deflection, the traditional device in optics to manipulate light, on its own cannot retain light-rays below the event horizon.

We arrange this article as follows.  In Section~\ref{sec:2} we explain the motions of light-rays in a gravitational field.  In Section~\ref{sec:3} we specify relations between the speed of light and gravity.  In Section~\ref{sec:4} we derive a differential version of Snell's law for black holes.  In Section~\ref{sec:5} we apply Snell's law to find the correct amount of light deflection by the Sun.  In Sections~\ref{sec:6} and \ref{sec:7} we study analytical and numerical trajectories of the rays.  In Section~\ref{sec:8} we discuss the Wheeler equation.  Brief discussions and conclusions are in Section~\ref{sec:9}.

\section{Huygens principle and gravity}\label{sec:2}
\begin{figure}
\resizebox{10cm}{!}{\includegraphics{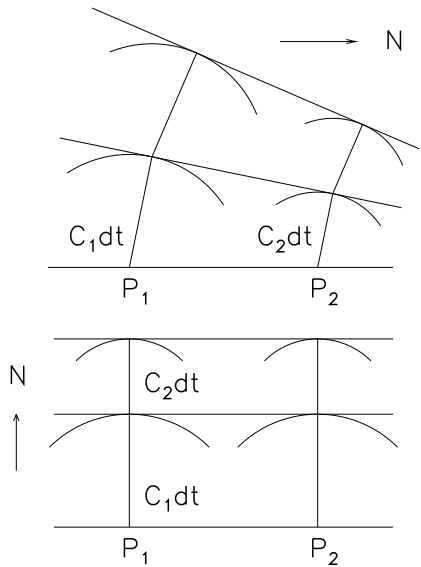}}
\caption{Illustrations of the Huygens principle originally from Einstein, where $P_1$ and $P_2$ are centres on the first wave front to emit spherical wavelets whose envelope becomes the next front to emit further wavelets, $C_1$ and $C_2$ are visual speeds of light (as defined in Section~\ref{sec:2}) which become slower the stronger the gravity, and $dt$ differential time of wavelet expansion.  The ray is launched either parallel to (upper part) or towards (lower part) the event horizon (location indicated by $N$).}
\end{figure}

In 1678 the Dutch physicist Huygens proposed that every point to which a luminous disturbance reaches becomes a source of a spherical wavelet, and the sum of these wavelets determines the form of the luminous disturbance at any subsequent time \cite{Huygens}.  He applied this principle to explain linear and spherical wave propagation, and to derive the laws of reflection and refraction.  In 1911 Einstein applied the same principle to explain deflection of light-rays in a gravitational field, and to outline the general theory of relativity \cite{Einstein}.  In FIG.~1 we redraw the historic illustration by Einstein with some modifications.  We follow Einstein to show how the wavelets emitted from two exemplary points on the first wave front, $P_1$ and $P_2$, expand and determine the second wave front.  In addition we show how the same process repeats itself to determine the third front.

In FIG.~1 the wave fronts can be seen as the floors and roof of a two level building, with the wavelets for pillars to support the first floor and roof.  Einstein stated: ``the velocity of light in the gravitational field is a function of the place" \cite{Einstein}.  From now on we will refer to Einstein's ``velocity of light" in the quotation as the speed of light seen by a distant observer, or the visual speed of light.  We also will refer to the velocity of light seen by a local observer the local speed of light, or simply the speed of light, which is a relativistic invariant.  In the upper part of the figure gravity is uneven at $P_1$ and $P_2$ so that the building tilts towards $P_2$ where the visual speed of light is slower due to stronger gravity.  In the lower part of FIG.~1 gravity is even at $P_1$ and $P_2$ so that the building does not tilt but with its levels piled closer to each other the stronger the gravity.

Now we have a problem: what would happen if a ray were launched from below the event horizon to approach the horizon at a right angle, as is shown in the lower part of FIG.~1?  It definitely will not make a U-turn and fall back to the centre, because it is well known the wavelets in the Huygens principle cannot propagate in the backward direction.  From the optical point of view, what could stop the ray from escaping across the event horizon?

We will show that light-rays launched outwardly from below the event horizon indeed encounter the horizon at right angles.  They are kept below the horizon not by deflection but by length contraction and time dilation, which in our view are new devices in the arsenal of optics to manipulate light, capable of lending deflection a helping hand to impede and eventually stop the escaping rays on their outward trajectories.

\section{visual speed of light and gravity}\label{sec:3}
Above the event horizon the Schwarzschild metric can be written as 
\begin{eqnarray}\label{eq:1}
(ds)^2 = (d\ell')^2 - c^2(dt')^2,\;\;\;\;\;r > r_s
\end{eqnarray}
where $c$ is the local speed of light, or speed of light, invariant under all circumstances,
\begin{eqnarray}\label{eq:2}
&&(d\ell')^2 = \bigg(1 - \frac{r_s}{r}\bigg)^{\!-1}\!\!\!(dr)^2 + r^2(d\Omega)^2,\nonumber\\
\\
&&(dt')^2 = \bigg(1 - \frac{r_s}{r}\bigg)(dt)^2\nonumber
\end{eqnarray}
measure length contraction and time dilation, $r_s$ being the Schwarzschild radius, $(d\Omega)^2 = (d\theta)^2 + \sin^2(\theta)(d\phi)^2$.  We are reminded the Schwarzschild metric is presented in the coordinate system $(t, r, \theta, \phi)$ accommodating a distant observer.  It is valid only in the presence of the factor $(1 - r_s/r)^{-1}$ for length contraction, and $1 - r_s/r$ for time dilation.  It tells us $d\ell'/dt' = c$, because $ds = 0$ for light-rays, that is the speed of light is invariant in the eyes of a local observer experiencing both length contraction and time dilation. But this is not true in the eyes of the distant observer.  Letting
\begin{eqnarray}\label{eq:3}
&&\cos^2(\alpha)\,(d\ell)^2 = (dr)^2,\nonumber\\
\\
&&\sin^2(\alpha)\,(d\ell)^2 = r^2(d\Omega)^2\nonumber
\end{eqnarray}
where $\alpha$ is the angle of incidence made by the ray and the radial direction,  explicitly specified as $\alpha_1$ and $\alpha_2$ at $B$ and $C$ in FIG.~2, we find from Eqs.~(\ref{eq:1}) nand (\ref{eq:2})
\begin{eqnarray}\label{eq:4}
\left(\frac{d\ell'}{dt'}\right)^{\!2} = \frac{r^2}{(r - r_s)^2}\bigg[1 - \frac{r_s}{r}\sin^2(\alpha)\bigg]\bigg(\frac{d\ell}{dt}\bigg)^{\!2}
\end{eqnarray}
or $(d\ell'/dt')^2 = V^2(d\ell/dt)^2$ to relate the local and visual speed of light.  We should take notice that in Eq.~(\ref{eq:4}) $d\ell'$ and $dt'$ are local measurements, whereas $r$, $r_s$, $\alpha$ and $d\ell$ are from the distant observer, comparable with say an object and its optical image.  Consequently $V$ can be compared with the refractive index that defines images in classical optics.  Indeed Eq.~(\ref{eq:4}) tells us clearly that, when $\alpha = 0$, we have $v = d\ell/dt = (1 - r/r_s)\,c$ which is the well known formula for the visual speed of a ray in the radial direction of Schwarzschild black hole from the eyes of the distant observer.

We follow the same procedure to investigate light-rays below the event horizon.  In this case the Schwarzschild metric is written as
\begin{eqnarray}\label{eq:5}
-(ds)^2 = (d\ell')^2 - c^2(dt')^2,\;\;\;\;\;r < r_s
\end{eqnarray}
where
\begin{eqnarray}\label{eq:6}
&&(d\ell')^2 = \bigg(\frac{r_s}{r} - 1\bigg)^{\!-1}\!\!\!(dr)^2 - r^2(d\Omega)^2,\nonumber\\
\\
&&(dt')^2 = \bigg(\frac{r_s}{r} - 1\bigg)(dt)^2\nonumber
\end{eqnarray}
which too leads to $d\ell'/dt' = c$ when $ds = 0$.  Eqs.~(\ref{eq:2}) and (\ref{eq:6}), like the Schwarzschild metric, are presented in the coordinate system $(t, r, \theta, \phi)$, valid only in the presence of the length contraction and time dilation factors. Eq.~(\ref{eq:4}) still applies but under the condition
\begin{eqnarray}\label{eq:7}
\sin^2(\alpha)\le\frac{r}{r_s}
\end{eqnarray}
in order to avoid $V$, defined below Eq.~(\ref{eq:4}), from becoming imaginary, a new issue arising from the term $-r^2(d\Omega)^2$ in Eq.~(\ref{eq:6}).

To understand Eqs.~(\ref{eq:4}) and (\ref{eq:7}) it is worth noting that Eqs.~(\ref{eq:2}) and (\ref{eq:6}) are our choice, or ansatz, to test if we are correct to anticipate light-rays below the event horizon.  Our choice assumes consistency between Eqs.~(\ref{eq:1}) and (\ref{eq:5}), true because they differ just by the sign of $(ds)^2$, which vanishes in the case of light-rays.  In addition our choice is evidently supported by the observation of Fiziev that the Wheeler equation has exact solutions both above and below the horizon \cite{Fiziev}.  In fact in the Wheeler equation the time factor, $\exp(i\omega t)$, (declared to be an ansatz by Fiziev) also is a choice to test if the equation has similar solutions both above and below the event horizon \cite{Fiziev, Wheeler}.  The test might fail, or become reduction to absurdity, in case the ansatz leads to a contradiction in general.  Since Eq.~(\ref{eq:6}) leads to a contradiction only when Eq.~(\ref{eq:7}) is violated, not in general, we might  be correct to expect light-rays below the event horizon, provided we can find the physical reason for Eq.~(\ref{eq:7}).

In the Schwarzschild metric length contraction takes place only in the $r$-direction.  Above the event horizon it has less of an effect in other directions due to the term $r^2(d\Omega)^2$ in Eq.~(\ref{eq:2}).  Below the event horizon, however, length contraction is always enhanced by the term $-r^2(d\Omega)^2$ in Eq.~(\ref{eq:6}).  When Eq.~(\ref{eq:7}) has been violated, the enhancement to length contraction turns out to be so significant that $(d\ell')^2$ in Eq.~(\ref{eq:6}) becomes negative, indicating the wavelets in the Huygens principle can no longer be in phase to direct their energy to form a thin pencil around a single direction, similar to what happens to the refractive ray when total reflection occurs.  Indeed the criterion for total reflection in classical optics does bear a fair resemblance to Eq.~(\ref{eq:7}).

Eq.~(\ref{eq:7}) defines a cone about the radial direction, with aperture $\rightarrow0$ when $r\rightarrow0$.  In a landmark paper in 1939 Oppenheimer and Snyder revealed that, by the gravitational deflection of light, all energy emitted from the surface of a collapsing star will almost all be reduced to within a cone about the outward normal, of progressively shrinking aperture as the star contracts \cite{Oppenheimer}.  From now on we will refer to the cone defined in Eq.~(\ref{eq:7}) as the internal Oppenheimer-Snyder cone, because it also is from gravitational deflection, located below the event horizon, aperture progressively shrinking with decreasing values of $r$, similar to what happens on the surface of a contracting star.

\vspace{1cm}
\section{Snell's law for black holes}\label{sec:4}

\begin{figure}[h]
\resizebox{6.67cm}{!}{\includegraphics{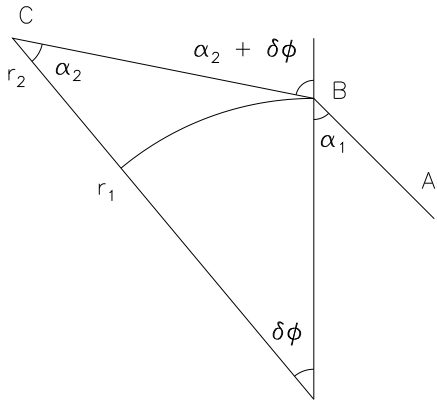}}
\caption{Passage of a ray defined as $r(\phi)$ in the $(r, \phi)$ plane of the spherical coordinates, $(r, \phi, \theta)$, with $r_1 = r(\phi_1)$, $r_2 = r(\phi_2)$, $\delta\phi = \phi_2 - \phi_1$.  The ray is shown to start from $A$, be deflected at $B$ with an angle of incidence, $\alpha_1$, and refraction, $\alpha_2 + \delta\phi$, both against the $r$-direction.  Eventually it reaches $C$ with an angle of incidence, $\alpha_2$, also measured against $r$.  The visual speed of the ray is determined by $V_1$ or $V_2$ when $r < r_1$ or otherwise.}
\end{figure}

We apply Eq.~(\ref{eq:4}) to specify the rule to determine the direction of light-rays in the Schwarzschild metric.  On account of the principle of phase matching we must have
\begin{eqnarray}\label{eq:8}
V_1\sin(\alpha_1) = V_2\sin(\alpha_2+\delta \phi)
\end{eqnarray}
for the ray in FIG.~2.  Here $V_1 = V(r_1, \alpha_1)$ and $V_2 = V(r_2, \alpha_2)$ are values of $V$ defined in Eq.~(\ref{eq:4}), $\alpha_1$ and $\alpha_2$ ray incidence against the outward normal, and $\delta\phi$ incremental angle made by $r$ when it rotates to take the values $r_1$ and $r_2$.  Eq.~(\ref{eq:8}) can be written as
\begin{eqnarray}
&&V_1\frac{\sin(\alpha_2 + \delta\phi) - \sin(\alpha_1)}{\alpha_2 + \delta\phi - \alpha_1}\,(\alpha_2 + \delta\phi - \alpha_1)\nonumber\\ 
&&= -\frac{V_2 - V_1}{\delta\phi}\sin(\alpha_2 + \delta\phi)\,\delta\phi\nonumber
\end{eqnarray}
which can readily be transformed into a differential equation.  To this end we divide the above equation with $\delta\phi$ and let $\delta\phi\rightarrow0$ so that $(\alpha_2 + \delta\phi - \alpha_1)/\delta\phi\rightarrow d\alpha/d\phi + 1$, assuming $\alpha_2 \rightarrow \alpha_1$ when $\delta\phi\rightarrow0$.  At the same time we find $\cos(\alpha_1)$ on the left hand side of the above equation.  We also find $dV/d\phi$ and $\sin(\alpha_1)$ on the right hand side of the equation, assuming $V_2\rightarrow V_1$ when $\delta \phi \rightarrow0$.  Letting $\alpha_1 = \alpha$ for generality,  we find via simple rearrangement of the terms that
\begin{eqnarray}
\frac{d\alpha}{d\phi}= -1 - \frac{r}{V}\frac{dV}{dr}\frac{\tan(\alpha)\,dr}{rd\phi}\nonumber
\end{eqnarray}
where
\begin{eqnarray}
\frac{\tan(\alpha)\,dr}{rd\phi}=\lim_{\delta\phi\rightarrow0}\frac{\tan(\alpha_2)\,(r_2 - r_1)}{r_1\delta\phi}=1\nonumber
\end{eqnarray}
from the geometry in FIG.~2, giving
\begin{eqnarray}\label{eq:9}
\frac{d\alpha}{d\phi} = -1-\frac{r}{V}\frac{dV}{d r},\;\;\;\;\;\;\frac{dr}{d\phi} = r\cot(\alpha)
\end{eqnarray}
as a differential version of Snell's law for Schwarzschild black holes.  With the help of the expression of $V$ in Eq.~(\ref{eq:4}), and some rather involved algebra, we find
\begin{eqnarray}\label{eq:10}
\left[1-\frac{2r_s}{r}\sin^2(\alpha)\right]\frac{d\alpha}{d\phi} = -\left[1 - \frac{r_s}{2r}\sin^2(\alpha)\right]\nonumber\\ 
+\bigg[1 - \frac{r_s}{r}\sin^2(\alpha)\bigg]\frac{r_s}{r - r_s}
\end{eqnarray}
to evaluate $d\alpha/d\phi$ in Eq.~(\ref{eq:9}).

\section{light deflection by the sun}\label{sec:5}
Letting $r_1 = 6.96\times10^8$ and  $r_s = 2.95\times10^3$ be the radius and Schwarzschild radius of the Sun in meters respectively, we have $r_s/r < r_s/r_1 = 4.24\times10^{-6}$ for a ray passing the Sun in a grazing angle.  Since in Eq.~(\ref{eq:4}) the term $(r_s/r)\sin^2(\alpha)$ represents a high order correction, due to inhomogeneous length contraction in different directions, we neglect similar terms in Eq.~(\ref{eq:10}) (radial gravity approximation) and find
\begin{eqnarray}\label{eq:11}
\frac{d\alpha}{d\phi}=-1 + \frac{r_s}{r - r_s}
\end{eqnarray}
where the effect of gravitational deflection is represented by the last term on the right hand side. To a very good approximation $r - r_s\simeq r_1/\cos(\phi)$ holds to give
\begin{eqnarray}\label{eq:12}
\alpha = \bigintsss_{-0.5\pi}^{\;0.5\pi}\!\!\!\!\frac{r_s}{r - r_s}\,d\phi\simeq\frac{2r_s}{r_1} = 1.75\mbox{ arc seconds}
\end{eqnarray}
as the amount of light deflection by the Sun, consistent with the well-known astronomical observation.

\section{analytical trajectories: radial gravity approximation}\label{sec:6}
\begin{figure}\label{fig:3}
\resizebox{9cm}{!}{\includegraphics{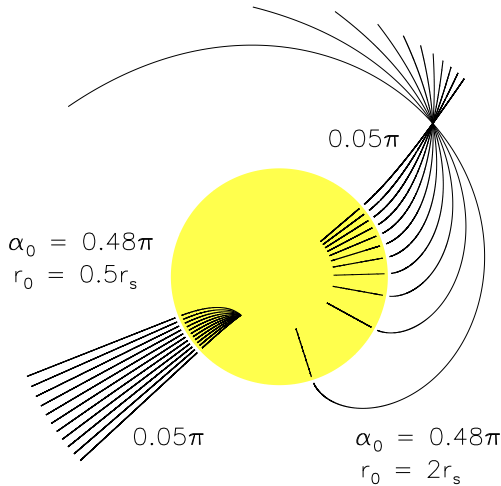}}
\caption{Event horizon (interior shaded, radius = $r_s$) and analytic trajectories of rays in radial gravity approximation from Eq.~(\ref{eq:14}) and detailed discussion in Section \ref{sec:6}.  The rays are launched at $r_0 = 0.5$ or $2r_s$ with $0.05\leq\alpha_0\leq0.48\pi$ until $r\sim r_s$.  Afterwards the rays are illustrated to continue their journeys along straight lines, leaving a schematic gap behind indicating length contraction and time dilation.}
\end{figure}

To illustrate the physics we integrate Eq.~(\ref{eq:11}) analytically for simplicity and clarity.  By using the second formula in Eq.~(\ref{eq:9}) to eliminate $d\phi$ we find
\begin{eqnarray}\label{eq:13}
\frac{r}{dr}\;\frac{d\sin(\alpha)}{\sin(\alpha)} = -1 + \frac{r_s}{r - r_s}
\end{eqnarray}
which is a closed relationship between $\alpha$ and $r$ ready to be integrated. The result is
\begin{eqnarray}\label{eq:14}
\frac{r/r_s}{1 - r_s/r}\sin(\alpha) = \frac{r_0/r_s}{1 - r_s/r_0}\sin(\alpha_0)
\end{eqnarray}
for trajectories of light-rays in radial gravity approximation, where $r_0$ and $\alpha_0$ are the initial conditions of the ray. Letting $r$ and $r_s$ be the radius and Schwarzschild radius of the Earth for example, we have $r/r_s\simeq r_0/r_s = 7.18\times10^8$ and Eq.~(\ref{eq:14}) tells us $\sin(\alpha)\simeq\sin(\alpha_0)$, that is a ray around us must travel along a straight path in its initial direction.  If the ray is launched to ascend and pulled back by gravity afterwards, then $\alpha = \pi/2$ must occur at some stage, giving via Eq.~(\ref{eq:14}) 
\begin{eqnarray}\label{eq:15}
\frac{r_s -r_0}{r_0^2}\,r^2 + \sin(\alpha_0)\,r - \sin(\alpha_0)\,r_s = 0
\end{eqnarray}
to determine the maximum value of $r$ the ray can reach.  Eq.~(\ref{eq:15}) has no real solution if
\begin{eqnarray}\label{eq:16}
\sin(\alpha_0)\leq2\,\sqrt{\;\,\frac{r_s}{r_0} - \bigg(\frac{r_s}{r_0}\bigg)^{\!2}}
\end{eqnarray}
which means the initial angle is too steep for the ray to be pulled back.  Eq.~(\ref{eq:16}) can be compared with Eq.~(\ref{eq:7}) and can be seen as an extended criteria for total reflection.  From now on we will refer to the cone defined by Eq.~(\ref{eq:16}) as the external Oppenheimer-Snyder cone, because the two authors discussed a similar cone above the surface of a collapsing star \cite{Oppenheimer}.

Eq.~(\ref{eq:16}) tells us that $\alpha_0 = 0$ when $r_0 = r_s$, that is, immediately above the event horizon, a ray still can defy gravity and escapt if it is launched in the radial direction.  On the other hand we have $\alpha_0 = \pi/2$ when $r_0 = 2r_s$ where lies the so-called phonon sphere \cite{Nitta} made from circular orbits of rays launched horizontally.

Data from Eq.~(\ref{eq:14}) are ready to be presented graphically to illustrate the physics.  All we have to do is to find $\phi$ from the second equation in Eq.~(\ref{eq:9}) with a simple numerical procedure.  To this end we need $\phi_0$, in addition to $r_0$ and $\alpha_0$, for the initial conditions to launch the rays.  Above the event horizon in FIG.~3 we launch rays at $r_0 = 2r_s$ and $\phi_0 = 0.25\pi$ (north-east direction) with $0.05\leq\alpha_0\leq 0.48\pi$ to descend and ascend in the clockwise and anticlockwise directions respectively. The ascending rays are traced until $r = 2.5r_s$ whereas the descending rays are traced to encounter the event horizon ($r\simeq r_s$ with $\alpha = 0$).  Afterwards the rays are shown to continue their journey to illustrate their straight trajectories ($\alpha\equiv0$) until $r = 0.5r_s$. 

We also launch rays below the event horizon at $r_0 = 0.5r_s$ and $\phi_0 = 1.25\pi$ (south-west direction) with $0.05\leq\alpha_0\leq0.48\pi$ to ascend in anticlockwise directions.  The rays are traced to encounter the event horizon ($r\simeq r_s$ with $\alpha = 0$) and continue their journey along straight lines ($\alpha\equiv0$) until $r = 2.5r_s$. We are reminded that $r = r_s$ is a singular point where $\alpha$ in Eq.~(\ref{eq:14}) has no definition.  Each of the trajectories in FIG.~3 is made from two pieces separated by a gap to indicate infinite length contraction and time dilation.

\section{numerical trajectories}\label{sec:7}

\begin{figure}\label{fig:4}
\resizebox{8cm}{!}{\includegraphics{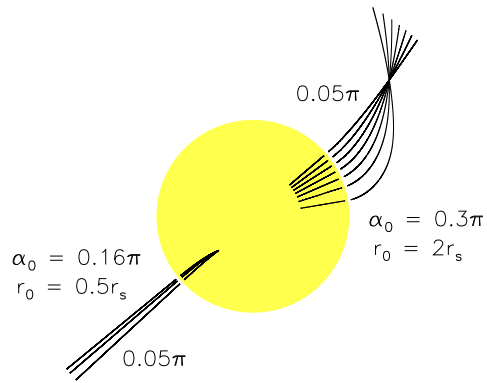}}
\caption{Numerical trajectories of light-rays from Eq.~(\ref{eq:17}) and detailed discussion in Section \ref{sec:7}, with conventions in FIG.~3, the schematic gaps indicating infinite length contraction and time dilation in particular.}
\end{figure}

To solve Eq.~(\ref{eq:10}) we use the second equation in Eq.~(\ref{eq:9}) to replace $d\phi$ with $d\alpha$ and apply the following numerical scheme
\begin{eqnarray}\label{eq:17}
\frac{u_{n + 1}}{u_n} = 1 + \frac{\Delta \xi_n}{\xi_n - 2u_n^2}\left[\frac{2 - \xi_n}{\xi_n - 1}\;\;\;\;\;\;\;\;\;\right.\nonumber\\ 
+ \left.\left(\frac{3}{2\xi_n} - \frac{1}{\xi_n - 1}\right)u_n^2\right]
\end{eqnarray}
where $u_n = \sin(\alpha_n)$, $\xi_n = r_n/r_s$, $\Delta\xi_n = \xi_{n + 1} - \xi_n$, $n = 0, 1, 2, ...$, iteratively for rays above the event horizon, with the initial condition $\xi_0 = 2$ and $\phi_0 = 0.25\pi$, $0.05\leq\alpha_0\leq0.3\pi$, and let the ray ascend until $\xi = 2.5$, or descend until $\xi = 1 + \Delta\xi$, that is one numerical step above the event horizon. Afterwards we let the rays continue their journeys along straight trajectories until $\xi = 0.5$.  We also apply Eq.~(\ref{eq:17}) with the initial condition $\xi_0 = 0.5$ and $\phi_0 = 1.25\pi$, $0.05\leq\alpha_0\leq0.16\pi$, and let the ray ascend until $\xi = 1 - \Delta\xi$, that is just one numerical step short below the event horizon.  Afterwards we let the rays continue their journeys along straight trajectories until $\xi = 2.5$.

We evaluate Eq.~(\ref{eq:17}) numerically with a first-order Runge-Kutta procedure written in house, always with 8000 steps when $\xi$ varies between 0.5 and 2.5. We use the analytical solution in Eq.~(\ref{eq:14}) as a reference to test accuracy of the outcome of Eq.~(\ref{eq:17}), with terms in $\sin^2(\alpha)$ removed from Eq.~(\ref{eq:10}).  We find that, over the range of $\xi$, the r.m.s.~deviation in $\alpha$ is $1.15\times10^{-4}$ on average when $\xi_0 = 0.5$ and $1.38\times10^{-4}$ when $\xi_0 = 2$.

Our numerical trajectories from Eq.~(\ref{eq:17}) are shown graphically in FIG.~4 ready to be compared with the analytical trajectories in FIG.~3.  We find that, if we do not assume $\sin(\alpha) = 0$ for radial gravity approximation, then we have to let $\alpha\le0.16\pi$ to restrict the initial incidence when the ray is launched at $\xi = 0.5$.  The reason apparently lies in the internal Oppenheimer cone in Eq.~(\ref{eq:7}) demanding $\alpha < 0.25\pi$ when $\xi = 0.5$.  Above the event horizon we find we have to let $\alpha\le 0.3\pi$ initially when the ray is launched at $\xi = 2$.

It is remarkable that, in FIG.~3, the ray launched with $r_0 = 2r_s$ and $\alpha_0 = 0.48\pi$ follows a nearly circular trajectory to orbit the event horizon.  Furthermore, with $r_0 = 2r_s$ and $\alpha_0 = \pi/2$, we can balance Eq.~(\ref{eq:14}) with $r\equiv2r_0$ and $\alpha\equiv\pi/2$, which represents a perfect circular trajectory.  However it is clear from FIG.~4 that, numerically, we cannot launch a ray to follow this trajectory.  The reason lies in that, if we let $\sin^2(\alpha) = 1$ in Eq.~(\ref{eq:10}), Eq.~(\ref{eq:13}) will be replaced by
\begin{eqnarray}
\frac{r}{dr}\;\frac{d\sin(\alpha)}{\sin(\alpha)} = -1 + \frac{r_s}{r - r_s} + \frac{r_s}{r - 2r_s}\nonumber
\end{eqnarray}
for transverse gravity approximation.  It tells us $r\equiv 2r_s$ and $\alpha\equiv\pi/2$ definitely cannot be a solution to Eq.~(\ref{eq:10}).  The phonon sphere, specified in Section~\ref{sec:4} as made from circular rays at $r = 2r_s$ in radial gravity approximation, cannot actually accommodate rays.  It is interesting that, according to Nitta, Chiba and Sugiyama, phonon orbits become unstable inside the phonon sphere \cite{Nitta}.

\section{Wheeler equation}\label{sec:8}
\begin{figure}\label{fig:5}
\resizebox{8cm}{!}{\includegraphics{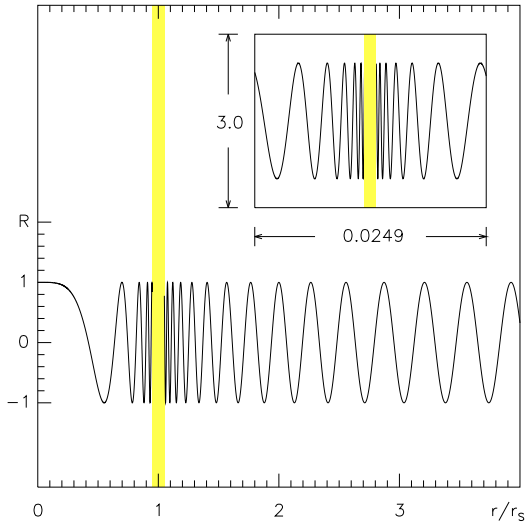}}
\caption{Real part of the solution to the Wheeler equation ($L = 0$, $k = 12.57/r_s$) in Eq.~(\ref{eq:18}) shown as a function of $r$.  A section of the solution, over a shaded range of 0.0249 in $r/r_s$ in the main frame, is shown in the inset for clarity.  A section of the solution in the inset is also shaded for clarity.  There can be an infinite number of such zoom-in procedures, like a Russian doll with infinite layers.  No definite boundary value can be imposed on the event horizon for the Fourier expansion over $[\,0,\,r_s\,]$.}
\end{figure}

For the Schwarzschild metric and weak electromagnetic waves, with negligible effect on spacetime curvature, the Wheeler equation assumes the form
\begin{eqnarray}\label{eq:18}
\frac{d^2R}{dx^{2}} + \left[k^2 -\frac{L(L + 1)}{r^2}\,e^\nu\right]R = 0 
\end{eqnarray}
where $k = \omega/c$, $\omega$ is the circular frequency of the waves, $L$ angular momentum, $e^\nu = 1 - r_s/r$ and $x = r + r_s\ln|r/r_s - 1|$ the so-called tortoise coordinate \cite{Fiziev}. If we add $3(r_s/r^3)\,e^\nu$ to the terms in the square brackets in Eq.~(\ref{eq:18}) then we find the Regge-Wheeler equation for gravitational waves which, unlike the Wheeler equation, is perturbational and approximate \cite{Regge}.

Eq.~(\ref{eq:18}) is reduced to the Riccati-Bessel equation in flat spacetime, where $r_s\rightarrow0$, $x\rightarrow r$ and $e^\nu\rightarrow1$.  If we let $R(r) = rZ(r)$ then Eq.~(\ref{eq:18}) is further reduced to the standard spherical Bessel equation in $Z(r)$ and its first and second order derivatives in $r$ \cite{Abramowitz}.  In curved spacetime, however, it becomes very challenging to study Eq.~(\ref{eq:18}), partly due to the simultaneous presence of both $x$ and $r$.  Fiziev found Eq.~(\ref{eq:18}) can be transformed into Heun's equation, with an explicit transform to remove $x$.  He also found a number of exact solutions, mostly in the form of power series \cite{Fiziev}.  Now the challenge lies in understanding physical implications of these solutions.  For example Fiziev was not certain if a complete set can be found in a functional space to present a natural basis of normal modes for an expansion of the solution to Eq.~(\ref{eq:18}) below the evwnt horizon \cite{Fiziev}. 

It is interesting that when $L = 0$ for electromagnetic disturbances without angular momentum, that is light-rays propagating in the radial direction, the presence of $r$ in Eq.~(\ref{eq:18}) is automatically removed, and the equation is reduced to the Helmholtz equation, with a simple solution in the tortoise coordinate:
\begin{eqnarray}\label{eq:19}
R(x)\propto\exp(i\omega t)\exp(-ikx)
\end{eqnarray}
of which the real part is evaluated and shown in FIG.~5.  It represents a harmonic wave propagating from $r = 0$ to $\infty$.  Clearly we are justified to expect the void below the event horizon being filled with electromagnetic disturbances, that serves as the premise of our analysis.

In Eq.~(\ref{eq:19}) we have $\omega t - kx = \mbox{const.}$ for the location of a wave front, giving $\omega dt  - kdx = \omega dt - kdr/(1 - r_s/r) = 0$, or $dr/dt = (1 - r_s/r)\,c$ as the speed of the wave.  Letting $\alpha = 0$ in Eqs.~(\ref{eq:2}) and (\ref{eq:3}) we also find $dr/dt = d\ell/dt = (1 - r_s/r)d\ell'/dt' = (1 - r_s/r)\,c$.  Apparently solutions to the Wheeler equation can be identified as light-rays in our black hole optics, at least in the radial direction, for the very reason why solutions to the Maxwell equation were identified as light-rays first time in history.

In Eq.~(\ref{eq:19}) the wave propagates outwardly, on account of the term $\exp(-ikx)$.  The wave can also propagate inwardly, because we can replace $\exp(-ikx)$ with $\exp(ikx)$.  This endorses our understanding that gravity cannot force a ray to make a U-turn and pull it back to the centre, if the ray is launched outwardly towards the event horizon at a right angle.  It must be time dilation and length contraction that impedes and eventually stops the escaping ray on its outward trajectory.

The effect of length contraction is clearly appreciable from FIG.~5.  The real part of $R(r)$ oscillates, pitch progressively higher as $r\rightarrow r_s$ indicating that the wave fronts of $R(r)$ are squeezed tighter the closer the distance from the event horizon.  Indeed, in the lower part of FIG.~1, the wave fronts from the Huygens principle are squeezed tighter towards the event horizon.  

In FIG.~5 we have to shade sections of $r$ close to the event horizon to hide high pitches for clarity, indicating $R(r)$ has no definite value at $r = r_s$ (that is $x = \infty$).  Therefore it may not be realistic to expect for a well-defined expansion of electromagnetic disturbances in the interior of the black hole, in terms of normal modes (solutions to the Wheeler equation with a definite boundary value at $r = r_s$ due to some physics) \cite{Fiziev}.

\section{conclusions}\label{sec:9}
The theory for the gravitational and electromagnetic fields is represented in an elegant set of compact equations, namely those of Einstein-Maxwell \cite{Bonnor, Lovelock, Papapetrou, Linet, Bachelot, Bachelot2, Gutsunaev, Choquet, Kroon, Blue, Ovsiyuk, Fernandes}, or the Wheeler equation coupled with the the Schwarzschild metric when spherical symmetry is assumed \cite{Wheeler}.  With an explicit substitution Fiziev transformed the Wheeler equation into Heun's equation which is known to have exact solutions both above and below the event horizon, in the form of spherical waves propagating both inwardly and outwardly \cite{Fiziev}.  Therefore the void below the event horizon may not be empty but filled with electromagnetic disturbances (light-rays) constantly trying to cross the horizon from below and escape.

To pull back or to slow down to a standstill?  That is the question when the task is to prevent a ray from crossing a certain barrier.  It turns out that, from the optical point of view, light-rays may not be kept below the event horizon of a black hole by deflection, which cannot force the rays to make a U-turn and fall back when they are propagating towards the horizon to escape.  It must be length contraction and time dilation that lend deflection a helping hand and slow the escaping rays to a standstill.

The optical point of view is in accordance with Einstein's vision with respect to light-rays in a gravitational field, in terms of the Huygens principle, when he outlined his plan to develop the general theory of relativity \cite{Einstein}.  It is immediately clear from his vision that deflection alone cannot stop light-rays from escaping across the event horizon, because the wavelets in the Huygens principle cannot propagate in the backward direction.  His vision also accommodates the possibility that light-rays may be slowed down to a standstill in a strong gravitational field, although at that time the concept of black hole has not yet been envisaged.

Oppenheimer and Snyder listed three reasons to explain why a collapsing star may close itself off from any communication with a distant observer, apart from via its gravitational field.  Two of these, the Doppler effect and gravitational red-shift, are devices from general relativity due to length contraction and time dilation.  The remaining reason, concerning the aperture of the external Oppenheimer-Snyder cone in Section~\ref{sec:6}, is due to gravitational deflection \cite{Oppenheimer}.  It certainly helps to stop light-rays from escaping, but may be compromised by the tendency of the light-rays from below the event horizon to encounter the horizon at right angles.

Our black hole optics interprets and verifies the observation of Fiziev that the Wheeler equation has exact solutions both above and below the event horizon, in the form of waves propagating in both the outward and inward directions.  Fiziev looks forward to identifying some physics, to impose a boundary condition upon the event horizon, that would enable him to find a natural basis of normal modes for a well-defined expansion of any small perturbation of the metric in the interior of the black hole \cite{Fiziev}.  Our black hole optics could become a useful alternative to help tackle this problem, because it entirely is based on the Schwarzschild metric, with no need for additional physics.

\end{document}